\documentclass{aip-cp}

\usepackage[numbers]{natbib}
\usepackage{rotating}
\usepackage{graphicx}
\usepackage{bm}
\usepackage{multirow}
\usepackage{epstopdf}

\begin{document}

\title{Comparative Study on Several Criteria for Non-equilibrium Phase Separation}

\author[aff1,aff2]{Yudong Zhang}
\author[aff1,aff3]{Aiguo Xu\corref{cor1} }
\author[aff1]{Guangcai Zhang }
\author[aff2]{Zhihua Chen }

\affil[aff1]{National Key Laboratory of Computational Physics, Institute of Applied Physics
and Computational Mathematics, P. O. Box 8009-26, Beijing, China.}
\affil[aff2]{Key Laboratory of Transient Physics, Nanjing University of Science and Technology, Nanjing 210094, China.}
\affil[aff3]{Center for Applied Physics and Technology, MOE Key Center for High Energy Density
Physics Simulations, College of Engineering, Peking University, Beijing 100871, China.}
\corresp[cor1]{Corresponding author: Xu\_Aiguo@iapcm.ac.cn}

\maketitle

\begin{abstract}
 Several different kinds of criteria for non-equilibrium phase separation to discriminate the two stages, the
 spinnodal decompostion (SD) and domain growth (DG), are compared and further investigated. The characteristic domain size and morphological function present two geometric criteria. Both of them can only provide rough estimations for the crossover from SD to DG. The reason for domain size is that the crossover in this description covers a process, instead of a specific time. The reason for the morphological function is that the result may rely on chosen threshold value. However, both the non-equilibrium strength and the entropy production rate are physical criteria and are more convenient to provide critical times. In fact, not only the non-equilibrium strength defined in the moment space opened by all the independent components of the used non-equilibrium quantities but also those defined in its subspaces can be used as criteria. Each of those criteria characterizes the phase separation process from its own perspective. Consequently, the obtained critical times may show slight differences. It should be pointed out that these slight differences are not contradictive, but consistent with each other and complementary in describing the complex phenomena.

\end{abstract}

\section{INTRODUCTION}
Phase separation is one of the most fundamental physical phenomena and is ubiquitous in industrial processes \cite{Sperling2006,PhaseTrans-book,Xu2015Progress}. Typical examples are crystal growth,  petroleum extraction, material processing and synthesis, etc. Understanding of the phase separation process is not only a fundamental scientific problem but also crucial for technological development. However, the hydrodynamics and kinetics of multiphase flows are much more complex, especially when it involves phase transition \cite{Xu2015Progress,Gan2015Discrete}. In recent years, great efforts have been made to investigate the phase separation by using experimental, theoretical, and numerical methods \cite{Ye2013Polymer,Yeganeh2014Anomalous,Iwashita2006Self}, among which the numerical method possesses a higher economy and flexibility and has been widely used in the research of phase separation \cite{Xu2015Progress,Succi2018The,Huang2015Multiphase,Gunstensen1991Lattice,Swift1996Lattice,Falcucci2010Lattice,Shan1993Lattice,Shan1994Simulation,Swift1995Lattice,He1999A,Li2016Lattice}.

Generally, under quenching condition, thermal phase separation process undergo two stages: the early spinodal decomposition (SD) and the late stage of domain growth (DG). The characteristics of the late stage have been extensively studied in the previous publications \cite{A1995Theory,Xu2006Morphologies}. In the DG stage, the characteristic domain size $R(t)$ grows exponentially with time $t$, i.e. $R(t)\sim t ^{\alpha}$ \cite{A1995Theory,Allen1976Mechanisms}. However, there is less research on the early stage of phase separation. In fact, how to exactly distinguish the SD stage from the DG stage is still an open problem. One way is to use the profile of characteristic domain size in the log-log scale. The moment when the power law of $R(t)$ appears can be regarded as the critical point of the two stages. However, it is difficult to get an accurate critical point by this method. In 2012, a new critical to distinguish two stages of phase separation was proposed based on the morphological method \cite{Gan2011Phase}. It was found that the boundary length $L$ increases at SD stage and decreases at DG stage, so the maximum point of $L$ can be used to mark the critical time. In 2015, with the help of non-equilibrium strength, another physical critical to distinguish the two stages of phase separation was provided \cite{Gan2015Discrete}. In our recent work, we find that the entropy production rate increases with time at the SD stage and decreases with time at the DG stage. The maximum point can also be used to indicate the critical time \cite{Zhang-2018arXiv2}.

So far, there are at least four kinds of criteria for thermal phase separation including characteristic domain size, morphological function, non-equilibrium strength, and the entropy production rate. Those methods provide the characteristics of phase separation process from different perspectives. In this work, we aim to investigate the differences and similarities between those different criteria by means of discrete Boltzmann method. The remainder of this paper is organized as follows. Section 2 introduces the multiphase model we used to simulate the thermal phase separation and the validation of the model. Section 3 shows and compares the evolution process of the phase separation from morphological function, characteristic domain size, non-equilibrium strength, and entropy production rate. Section 4 concludes the present paper.

\section{METHOD and VALIDATION}

\subsection{Discrete Boltzmann model for nonideal fluid}
As a kinetic modeling of non-equilibrium and complex flow, the discrete Boltzmann model (DBM) has been widely used in various flows including high speed compressible flow \cite{Gan2017Three}, flow instability \cite{Feng2016Viscosity,Lai2016Nonequilibrium,Lin2017Discrete}, combustion and detonation \cite{Lin2017A}, and non-equilibrium pahse transition \cite{Gan2015Discrete}. In order to describe the nonideal equation of state (EOS) and the surface tension effect, the collision term on the right side is corrected by adding a extra term then the evolution equation of discrete Boltzmann model reads \cite{ZYD-2018arXiv}
\begin{equation}\label{Eq-DBM1}
\frac{{\partial f_{ki}}}{{\partial t}} + {\bf{v}}_{ki} \cdot \nabla f_{ki} =  - \frac{1}{\tau }\left( {f_{ki} - {f_{ki}^s}} \right) + I_{ki},
\end{equation}
where $f_{ki}$ is distribution function of the discrete velocity ${\bf{v}}_{ki}$ and the subscript $ki$ denotes the index of discrete velocity. $f_{ki}^s$ is the discrete Shakhov distribution function which reads
\begin{equation}\label{Eq-fs}
{f_{ki}^s} = {f_{ki}^{eq}}\left\{ 1+ {(1 - \Pr ){\mathbf{c}}_{ki} \cdot {\mathbf{q} }\left[ {\frac{{{c_{ki}^2}}}{{RT}} - (D+2)} \right]/((D+2)\rho R^2T^2)} \right\},
\end{equation}
where $f_{ki}^{eq}$ is the discrete local equilibrium distribution function, $D$ represents space dimension, and $R$ indicates gas constant. $\mathbf{u}$ and $T$ are macro velocity and temperature, respectively. $\mathbf{q}$ represents heat flux, ${\mathbf{c}}_{ki}={\mathbf{v}}_{ki}-\mathbf{u}$, and $c_{ki}^2={\mathbf{c}}_{ki} \cdot {\mathbf{c}}_{ki}$. The discretization of particle velocity space and the calculation of $f_{ki}^{eq}$ are referred to Ref. \cite{ZYD-2018arXiv,Watari2003Two,Watari2016Is}. The Shakhov model rather than the BGK model is adopted to replace the original collision integral, which possesses an adjustable Prandtl number and have more advantages in investigating the non-equilibrium flows. The extra term $I_{ki}$ are used to describe the interparticle forces which is similar to the one introduced by Klimontovich for nonideal gases \cite{Gonnella2007Lattice,Klimontovich-book} and reads
\begin{equation}\label{Eq-Iki}
I_{ki} =  - \left( {{A_0} + {\mathbf{A}_{1}} \cdot {\mathbf{c}_{ki} } + {A_{2,0}}{c_{ki}^2}} \right)f_{ki}^{eq},
\end{equation}
where $A_0$, $\mathbf{A}_1$, and $A_{2,0}$ are three parameters. According to Chapman Enskog expansion, the Navier-Stokes equations can be derived from Eq. (\ref{Eq-DBM1}) without the extra term $I_{ki}$. Here, taking into account the extra term $I_{ki}$, we need to derive the hydrodynamic equations for nonideal fluid proposed by Onuki \cite{Onuki2005Dynamic,Onuki2007Dynamic}. Then the parameters $A_0$, $\mathbf{A}_1$, and $A_{2,0}$ can be calculated from the Champan-Enskog expansion which read
\begin{equation}\label{Eq-A0}
A_0 =  - 2 A_{2,0}T,
\end{equation}
\begin{equation}\label{Eq-A1}
{\bf{A}}_1 =  \frac{1}{{\rho T}} \nabla \cdot \left[ {\left( {{P} - \rho T} \right) {\bf{I}} + \bm{\Lambda}} \right],
\end{equation}
\begin{equation}\label{Eq-A20}
\begin{array}{l}
 A_{2,0}= \frac{1}{{2\rho {T^2}}}\left[ {\left( {{P} - \rho T} \right)\nabla  \cdot {\bf{u}} + {\bm{\Lambda }}:\nabla {\bf{u}} + a{\rho ^2}\nabla  \cdot {\bf{u}}} \right.\\
\left. {{\kern 10pt}  - K(\frac{1}{2}\nabla \rho  \cdot \nabla \rho \nabla  \cdot {\bf{u}} + \rho \nabla \rho  \cdot \nabla \left( {\nabla  \cdot {\bf{u}}} \right) + \nabla \rho  \cdot \nabla {\bf{u}} \cdot \nabla \rho )} \right].
\end{array}
\end{equation}
where $P = \frac{{\rho T}}{{1 - b \rho}} - a {\rho}^2$ is the van der Waals (vdW) pressure, ${\bf{I}}$ is the unit tensor, $\bm{\Lambda}  =  - \left[ {M\rho {\nabla ^2}\rho  + \frac{M}{2}{{\left| {\nabla \rho } \right|}^2}{\rm{ + }}\rho T\nabla \rho  \cdot \nabla \left( {M/T} \right)} \right]{\bf{I}} + M\nabla \rho \nabla \rho$ is the contribution of surface tension to the pressure tensor, and $M=K+CT$ with $K$ and $C$ are two constants. It can be seen that all of the parameters $A_0$, $\mathbf{A}_1$, and $A_{2,0}$ are only depend on the quantities $\rho$, $\bf{u}$, $T$ and their spatial derivative so they can be readily computed at each time step.

\subsection{Numerical verification}
In order to validate the DBM for nonideal fluid, the liquid-vapor coexistence points at various temperatures are simulated. The computational grids are $Nx \times Ny= 200 \times 4$ with space step $\Delta x =\Delta y =0.01$. Time step is $\Delta t=0.0001$ and relaxation time $\tau =0.02$. Periodic boundary conditions are adopted in both directions. The first and second order spatial derivatives are all calculated by the nine-point stencils (NPS) \cite{Tiribocchi2009Hybrid,Gan2012FFT} scheme, which possesses a higher isotropy and is able to reduce spurious velocities significantly. The time derivation is solved by the first order forward difference. The coefficient of surface tension is $K=5\times 10^{-5}$ when calculate the force term. The parameters $a$ and $b$ in the vdW EOS are chosen as $a=9/8$ and $b=1/3$ so it has a critical point at $\rho_c = T_c=1$.

\begin{figure}[h]
\centering
  \includegraphics[height=5cm]{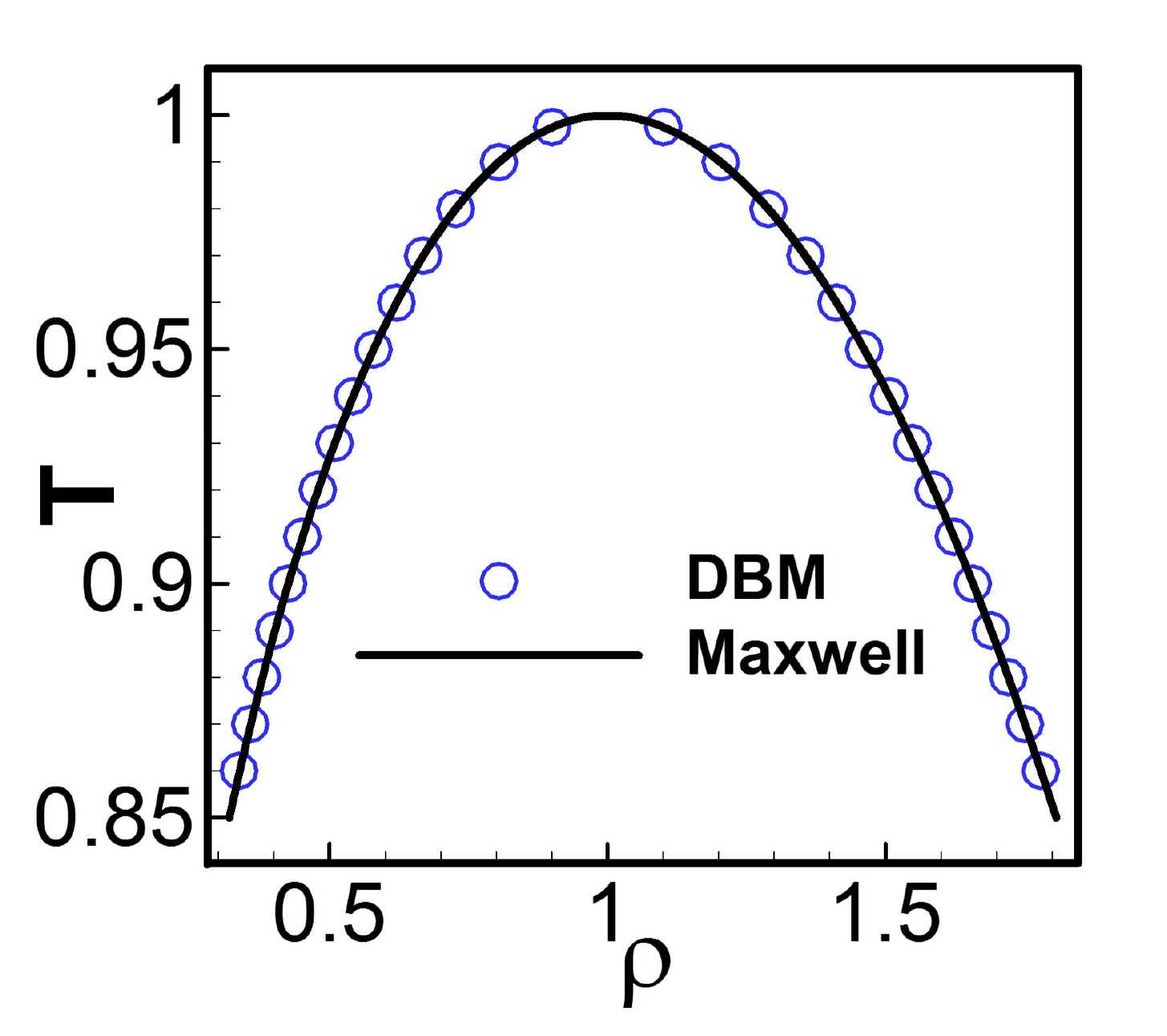}
  \caption{The liquid-vapor coexistence curve for a van der Waals fluid. The symbols represent DBM results and the solid line is the analytical solution calculated by Maxwell construction.}
  \label{fig1}
\end{figure}

The initial conditions are
\begin{equation}\label{Eq-Initial1}
\left\{ \begin{array}{l}
{(\rho ,T,{u_x},{u_y})_L} = ({\rho _v},0.9975,0,0),\\
{(\rho ,T,{u_x},{u_y})_M} = ({\rho _l},0.9975,0,0),\\
{(\rho ,T,{u_x},{u_y})_R} = ({\rho _v},0.9975,0,0),
\end{array} \right.
\end{equation}
where the subscript ``$L$'', ``$M$'', and ``$R$'' indicate the regions $x \leq \frac{1}{4}Nx$, $\frac{1}{4}Nx< x \leq \frac{3}{4}Nx$, and $x>\frac{3}{4}Nx$, respectively, $\rho _v =0.955$ and $\rho _l =1.045$ are theoretical vapor and liquid densities at $T=0.9975$. The temperature is dropped to $T=0.99$ when the equilibrium state of the system is achieved. Then the temperature drops by a small value $\Delta T=0.01$ each time once the equilibrium state of the system is achieved again. Simulations go on until the temperature is reduced to $0.85$ then a series of coexistence points are obtained. The results are shown in Fig. \ref{fig1}. The solid line is directly calculated from vdW EOS using a Maxwell construction. It shows that the coexistence points simulated by the DBM are well agree with the coexistence curve, which verifies that the multiphase DBM provides a correct vdW EOS.

\begin{figure}[h]
\centering
  \includegraphics[height=5cm]{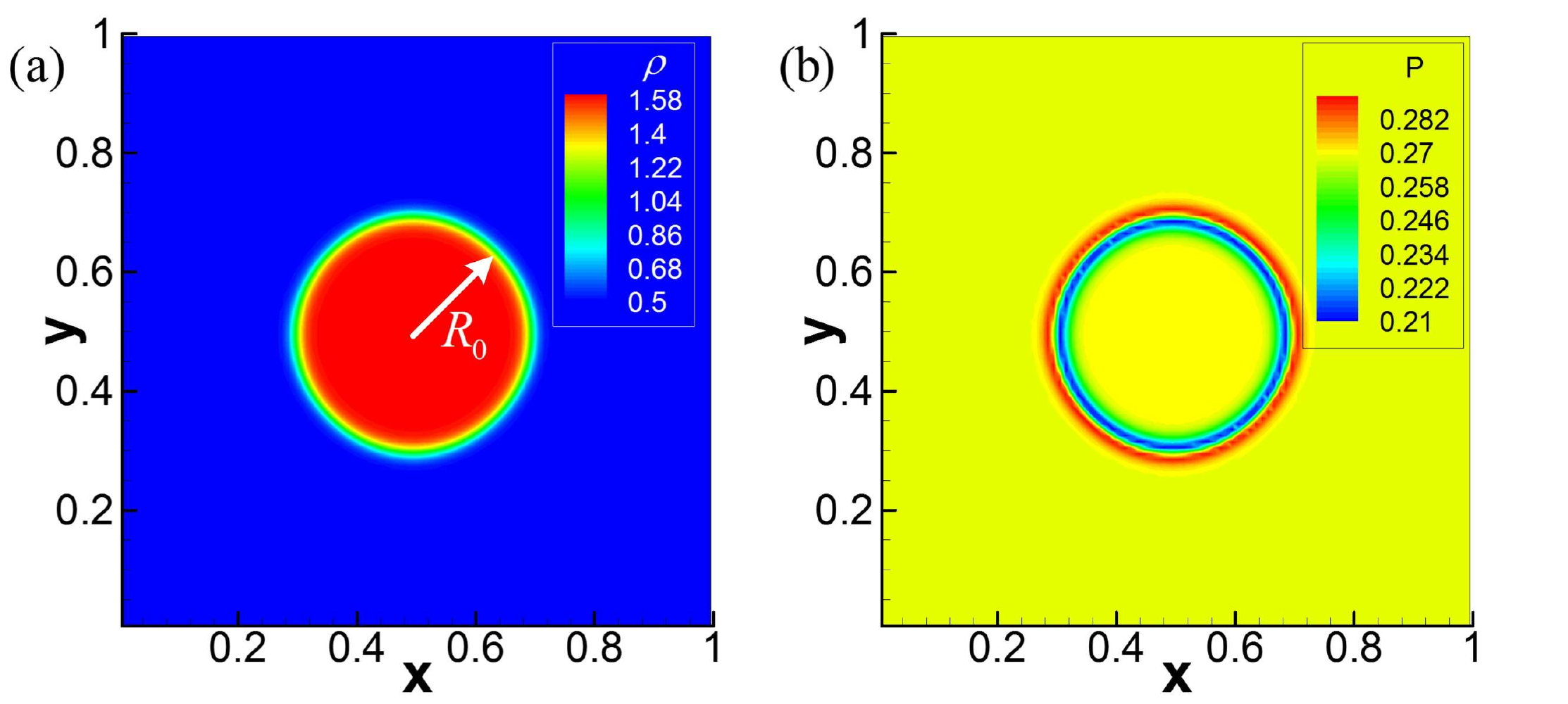}
  \caption{The contour maps of (a) density and (b) pressure for a circular droplet with a radius of $R_0$ at steady state.}
  \label{fig2}
\end{figure}

Then we check the surface tension of the new model by the Laplace's law. As shown in Fig. \ref{fig2}, a circular droplet with a radius of $R_0$ are surrounded by its vapor. The initial conditions are set as
\begin{equation}\label{Eq-Initial2}
\left\{ \begin{array}{l}
{(\rho ,T,{u_x},{u_y})_{{\rm{in}}}}{\kern 3pt} = (1.5865,0.92,0,0),\\
{(\rho ,T,{u_x},{u_y})_{{\rm{out}}}} = (0.4786,0.92,0,0),
\end{array} \right.
\end{equation}
where the subscript ``in'' and ``out'' indicate the regions inside and outside the circle, respectively. All the rest parameters are the same with those in Fig. \ref{fig1}. Periodic boundary conditions are adopted in both directions. The NPS scheme is used to calculate the spatial derivatives and the time derivation is solved by the first order forward difference.

\begin{figure}[h]
\centering
  \includegraphics[height=6cm]{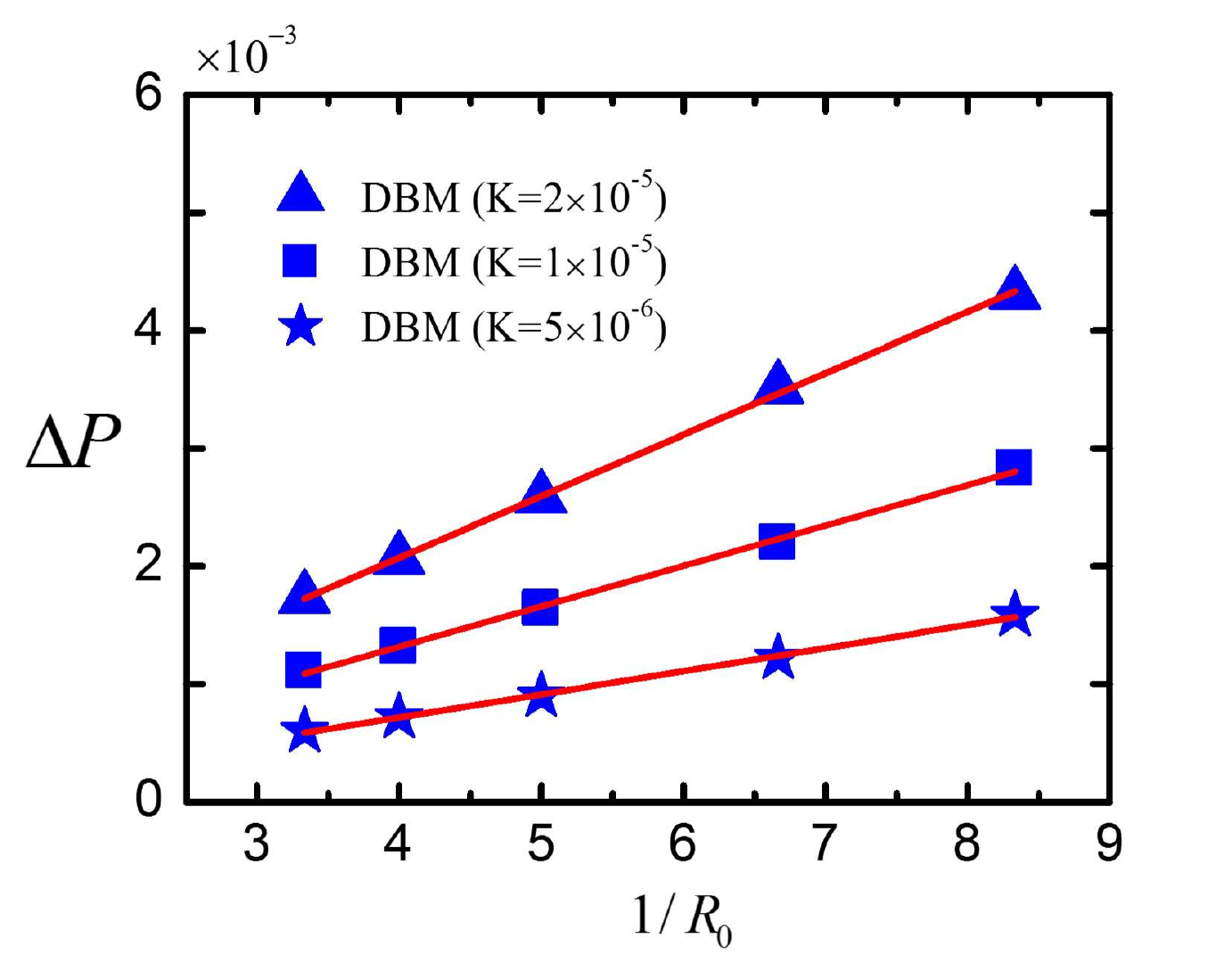}
  \caption{$\Delta P$ plotted vs $1/R_{0}$ for three different coefficients of surface tension as tests of Laplace's law. The symbols are DBM results while the lines are linear fits.}
  \label{fig3}
\end{figure}

According to Laplace law, the pressure difference $\Delta P$ between the inside and outside of the circular domain is proportional to the reciprocal of radius $1/R_0$ when the surface tension is fixed,
\begin{equation}\label{Eq-Laplace1}
\Delta P =\frac{\sigma }{R_0},
\end{equation}
where $\sigma$ is the surface tension. In the simulation, three different values of $\sigma$ are used by changing the coefficient $K$. Figure \ref{fig2} (a) and \ref{fig2} (b) shows the density and pressure contour at steady state, respectively. The pressure difference $\Delta P$ between the inside and outside of the circular domain under different circle radiuses and surface tensions are given in Fig. \ref{fig3}. The DBM results are denoted by symbols and the lines are obtained by linear fitting. There is a linear relationship between the pressure difference $\Delta P$ and the reciprocal of radius $1/R_0$. We can see that the results of DBM is consistent with the Laplace's law.

\section{RESULTS and ANALYSIS}
In this section, the thermal phase separation process is simulated. The initial conditions are
\begin{equation}\label{Eq-Initial-set}
(\rho, u_x, u_y, T)=(1+\delta, 0.0, 0.0, 0.9),
\end{equation}
where $\delta$ represents a random density noise with an amplitude of 0.01. The simulated area is $Lx \times Ly=1.0 \times 1.0$ with the space step $\Delta x =\Delta y =0.01$. Time step is $\Delta t= 1\times 10^{-4}$ and the relaxation time $\tau =0.02$. The parameters in the vdW EOS are chosen as $a=9/8$ and $b=1/3$, so the critical density $\rho _c$ and critical temperature $T_c$ are $\rho _c= T_c =1.0$. The coefficient of surface tension is $K=1 \times 10^{-5}$. The spatial derivations are calculated by NPS scheme and the time derivation is solved by the first order forward difference. Periodic boundary conditions are used in both directions.

\begin{figure}[h]
\centering
  \includegraphics[height=7cm]{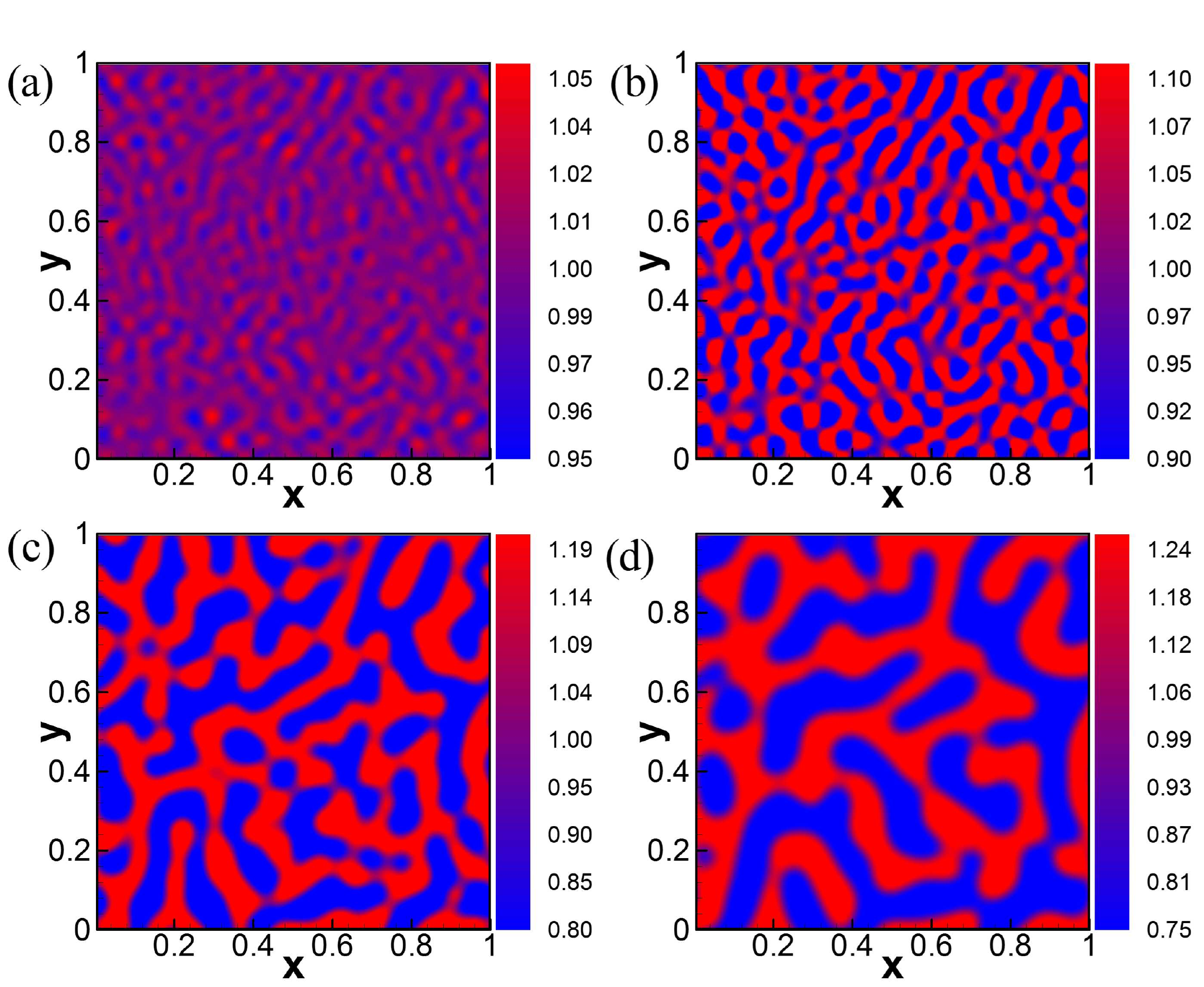}
  \caption{Density contour maps at several times for thermal phase separation. (a) $t=0.5$, (b) $t=1.0$, (c) $t=5.0$, (d) $t=10.0$.}
  \label{fig4}
\end{figure}

The evolution process of the thermal phase separation is shown in Fig. \ref{fig4}. The density contour at several typical times are plotted. Figure \ref{fig4} (a) corresponds to the time $t=0.5$ which is in the SD stage. We can see that the fluid separates into small regions but the interfaces are blurry at this stage. Figure \ref{fig4} (b) corresponds to the time $t=1.0$ which is around the critical time. Compared with Fig. \ref{fig4} (a), there is little change in the phase regions but the interface is much clearer. Figure \ref{fig4} (c) and \ref{fig4} (d) represent the SD stage at time $t=5.0$ and $t=10.0$, respectively. In this stage, small domains merge with each other and the phase regions grow quickly under the action of surface tension. From the density contour map, we can only get parts of the information of thermal phase separation qualitatively. In the following parts, we further conduct a quantitative analysis of the process of phase separation.

\subsection{Morphological characterization}
Morphology is a theory of analyzing spatial structure. It was first used to study the gas permeability of porous media then it was well developed and widely used in the image processing and analysis. In recent years, morphological analysis technology has also been applied in the study of the complex physical fields including reaction diffusion system \cite{Mecke1996Morphological}, dynamic response of porous materials under shock \cite{Xu2009Morphological,Xu2016Complex}, and the phase separation process of complex fluids \cite{Xu2006Morphologies,Sofonea1999Morphological}, etc. It is becoming an effective tool and plays an important role in data analysis and information extraction.

 In morphology, the Minkowski functionals can fully describe the geometric properties of a $D$-dimensional convex sets satisfying the morphological properties. A physical field can be described by $\Theta (\mathbf{r})$, where $\mathbf{r}$ is the position in a $D$-dimensional space and $\Theta$ is a physical variable such as density, temperature, and velocity. If we set a threshold value $\Theta_{th}$, the regions is defined as white when $\Theta (\mathbf{r}) \geq \Theta_{th}$ and black when $\Theta (\mathbf{r})< \Theta_{th}$, then a pattern of white and black pixels is obtained. The points with $\Theta (\mathbf{r})\geq \Theta_{th}$ compose the $D$-dimensional convex set and its morphological properties can be completely described by $D+1$ functionals. For the two-dimensional case, the three Minkowski functionals are the white area fraction $A$, the reduce total boundary length $L$, and the Euler characteristic $\chi$.

\begin{figure}[h]
\centering
  \includegraphics[height=6cm]{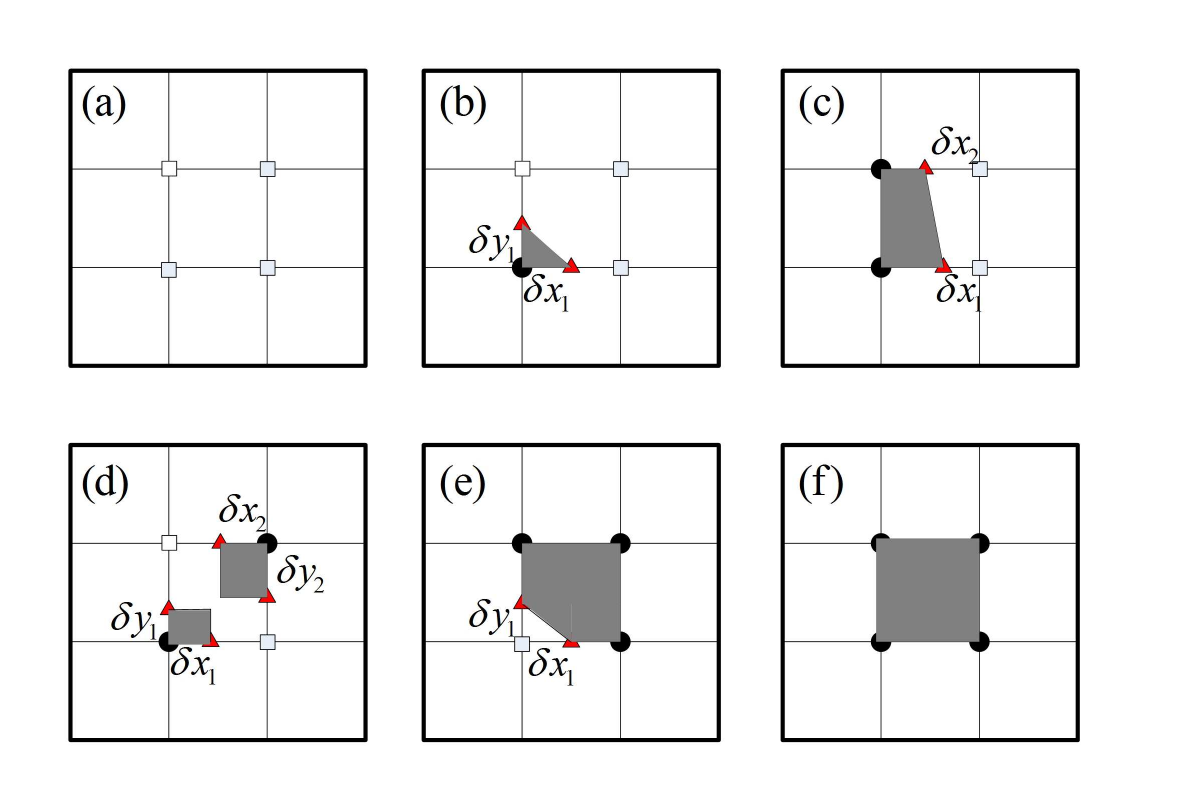}
  \caption{Several different kinds of pixels. The symbols of square, circle, and triangle denote the density at this point are above, below, and equal to the threshold value $\rho_{th}$, respectively. The positions of the triangle symbols are calculated by linear interpolation.}
  \label{fig0-1}
\end{figure}

For example, if we need to analyze the density contour map, we should first choose a density threshold value $\rho_{th}$ then the white area corresponds to the high density and the black area corresponds to the low density. The first Minkowski functional $A$ is the white area $A_w$ divided by the total area $A_{total}$,
\begin{equation}\label{Eq-A}
A=\frac{A_w}{A_{total}} .
\end{equation}
It should be noted that a pixel here corresponds to a small square formed by four grid nodes in the simulation results as shown in Fig. \ref{fig0-1}. The white area directly plus one if the values of density at the four vertices are all above $\rho _{th}$ as shown in Fig. \ref{fig0-1} (a), otherwise the white area plus a interpolated number between zero and one. Figs. \ref{fig0-1} (b)-(f) show several different kinds of pixels and the calculation of $\Delta A_w$ can be found below Fig. \ref{fig0-1}.  The ratio of the white area is gradually reduced from one to zero when $\rho_{th}$ increases from the lowest value to the highest one.

The second Minkowski functional $L$ is the length of the boundary between the white and black regions $L_w$ divided by the circumference of the simulated region $L_c$,
\begin{equation}\label{Eq-L}
L=\frac{L_w}{L_{c}} .
\end{equation}
With the increasing of $\rho_{th}$, the value of $L$ increase from zero then arrives its maximum value and finally decrease to zero again.

The third Minkowski functional $\chi$ is defined as
\begin{equation}\label{Eq-Chi}
\chi=n^w-n^b,
\end{equation}
where $n^w$ and $n^b$ are the number of simply connected white domains and black domains, respectively. The Euler characteristic $\chi$ describe the connectively of the domains in a purely topological way.

The operations for the different cases in Fig. \ref{fig0-1} are as follow,
\begin{itemize}
  \item Case (a): $A_w=A_w+1$.
  \item Case (b): $A_w=A_w+\Delta A_w$, $L_w=L_w+ \Delta L_w$, $\chi=\chi -1$, \\
  where $\Delta A_w=1-\frac{1}{2}\delta x_1 \delta y_1 $ and $\Delta L_w=\sqrt{\delta x_1 ^2+\delta y_1 ^2}$.
  \item Case (c): $A_w=A_w+\Delta A_w$, $L_w=L_w+ \Delta L_w$, \\
  where $\Delta A_w=\frac{1}{2}(\delta x_1 + \delta x_2 )$ and $\Delta L_w=\sqrt{(\delta x_2 - \delta x_1) ^2+1}$.
  \item Case (d): $A_w=A_w+\Delta A_w$, $L_w=L_w+ \Delta L_w$, $\chi=\chi + \Delta \chi$, \\
  where $\Delta L_w=\delta x_1+\delta y_1+\delta x_2+\delta y_2$,

  if $\Delta L_w < 2$, $\Delta A_w=1- (\delta x_1 \delta y_1+\delta x_2 \delta y_2)$ and $\Delta \chi =-2$;

  if $\Delta L_w = 2$, $\Delta A_w=0.5$ and $\Delta \chi =-2$;

  if $\Delta L_w> 2$, $\Delta A_w=(1-\delta x_1) (1-\delta y_1)+(1-\delta x_2) (1-\delta y_2)$ and $\Delta \chi =2$.
  \item Case (e): $A_w=A_w+\Delta A_w$, $L_w=L_w+ \Delta L_w$, $\chi=\chi +1$, \\
    where $\Delta A_w=\frac{1}{2}\delta x_1 \delta y_1 $ and $\Delta L_w=\sqrt{\delta x_1 ^2+\delta y_1 ^2}$.
  \item Case (f): no operation.
\end{itemize}

\begin{figure}[h]
\centering
  \includegraphics[height=8cm]{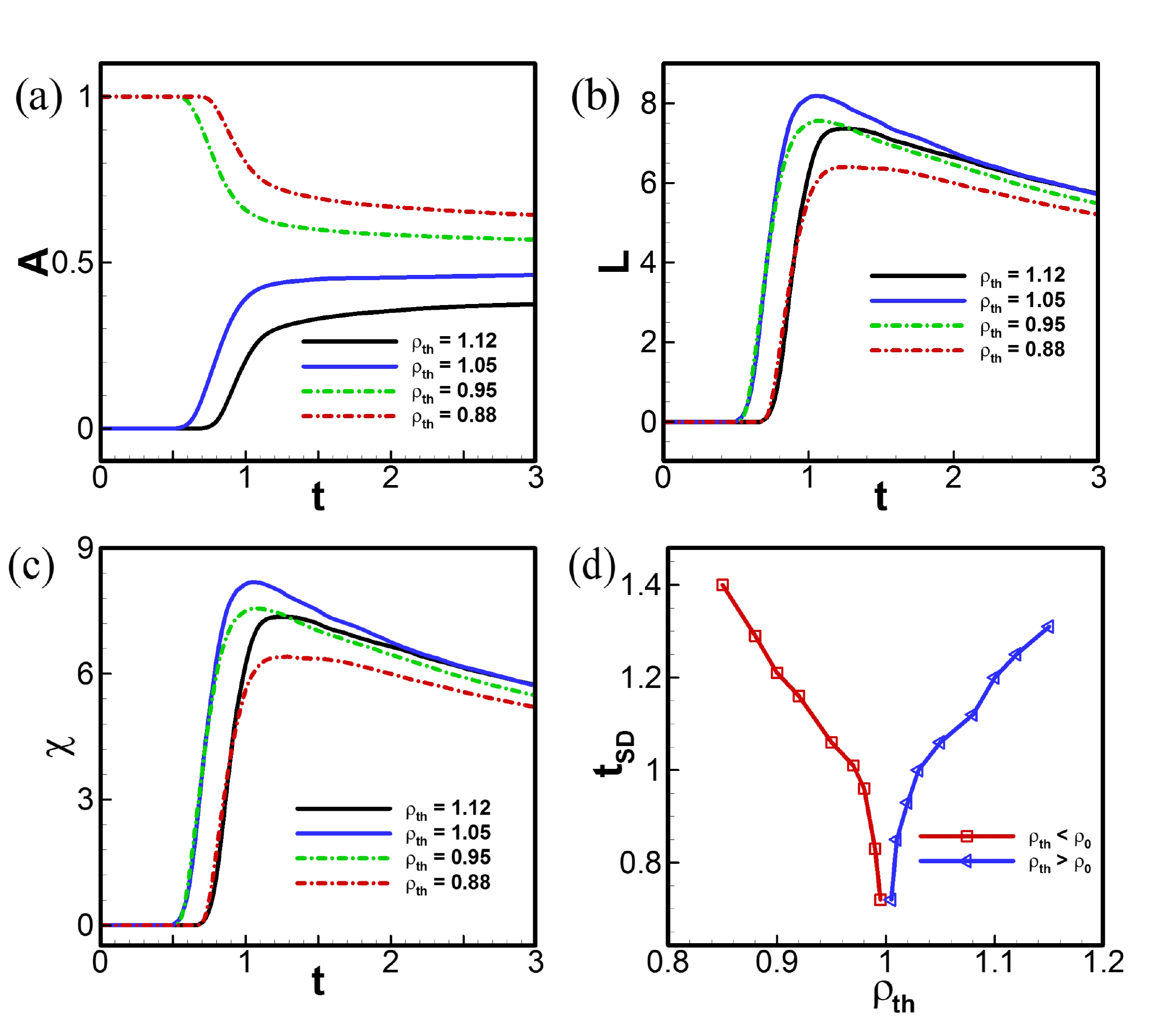}
  \caption{Morphological characteristics of the thermal phase separation. (a) Profiles of white area fraction $A$ with different threshold values $\rho _{th}$. (b)Profile of boundary length $L$ with different $\rho _{th}$. (c) Profiles of Euler characterisitc $\chi$ with different $\rho _{th}$. (d) Relationship between the critical time $t_{SD}$ and threshold value $\rho _{th}$.}
  \label{fig5}
\end{figure}

Morphological characteristics of the density contour map for thermal phase separation are shown in Fig. \ref{fig5}. The profiles of $A$, $L$, and $\chi$ are plotted in Fig. \ref{fig5} (a), Fig. \ref{fig5} (b), and Fig. \ref{fig5} (c), respectively. At the first stage of phase separation, due to the growth of density fluctuations, a large number of interfaces are buildup. This changes result in the increase of the boundary length $L$. However, at the DG stage, with the coalescence of small domains and the larger domains form, the Euler characteristic decreases and the total boundary length also decreases. Thus, the boundary length can be used as a geometric criterion to identify the two stage. The boundary length increases at the SD stage and decreases at the DG stage, so the maximum point corresponds to the critical time $t_{SD}$\cite{Gan2011Phase}.

However, it should be noted that the profile of boundary length $L$ depend on the threshold value $\rho _{th}$  we choose. Different profiles can be obtained by using different threshold values as shown in Fig. \ref{fig5} (b). As a result, the maximum point of the $L$ also depends on the threshold value. The relationship between the critical time $t_{SD}$ and the threshold value $\rho _{th}$ are given in Fig. \ref{fig5} (d). It can be seen that $t_{SD}$ decrease with the increase of the $\rho_{th}$ when $\rho_{th} < \rho_0$, where $\rho_0$ is the initial density without fluctuation. However, when $\rho_{th} < \rho_0$, $t_{SD}$ increase with the increase of the $\rho_{th}$. From this relationship, we learn that the criterion given by morphological method is not unique. The value of the  $t_{SD}$ depend on the threshold value we choose.

\subsection{Thermodynamic non-equilibrium strength and entropy production}
It has been known that the DBM can provide a convenient  measure for the thermodynamic non-equilibrium effects by means of the higher order kinetic moments. The non-equilibrium quantities are defined as the differences between the kinetic moments of $f$ and their corresponding local equilibrium distribution $f^{eq}$ , which read \cite{Xu2012Lattice}
\begin{equation}\label{Eq-Deltaxing-m}
{\bm{\Delta }}_m^* = {\bf{M}}_m^*\left( {{f_{ki}}} \right) - {\bf{M}}_m^*\left( {f_{ki}^{eq}} \right),
\end{equation}
where ${\bf{M}}_m^*\left( f \right)$ indicates the $m$th order kinetic center moment,
\begin{equation}\label{Eq-Mxing-m}
{\bf{M}}_m^*\left( f \right){\rm{ = }}\sum\limits_{ki} {{f_{ki}}\underbrace {{({\bf{v}}_{ki}-{\bf{u})}} ({\bf{v}}_{ki}-\mathbf{u}) \cdots {({\bf{v}}_{ki}-\mathbf{u})}}_m}.
\end{equation}

Recently, we find that the non-equilibrium strength $D_m^*$ is more suitable for characterizing the fluid interface \cite{ZYD-2018arXiv}. The definitions of the first four non-equilibrium strengths are rewritten here
\begin{equation}\label{D2}
D^*_2 = \sqrt {{{(\Delta _{2,xx}^*)}^2} + {{(\Delta _{2,xy}^*)}^2} + {{(\Delta _{2,yy}^*)}^2}},
\end{equation}
\begin{equation}\label{D31}
D^*_{3,1} = \sqrt {{{(\Delta _{3,1,x}^*)}^2} + {{(\Delta _{3,1,y}^*)}^2}},
\end{equation}
\begin{equation}\label{D3}
D^*_{3} = \sqrt {{{(\Delta _{3,xxx}^*)}^2} + {{(\Delta _{3,xxy}^*)}^2}+ {{(\Delta _{3,xyy}^*)}^2}+ {{(\Delta _{3,yyy}^*)}^2}},
\end{equation}
\begin{equation}\label{D42}
D^*_{4,2} = \sqrt {{{(\Delta _{4,2,xx}^*)}^2} + {{(\Delta _{4,2,xy}^*)}^2} + {{(\Delta _{4,2,yy}^*)}^2}}.
\end{equation}
Then we can further define the total non-equilibrium strength $D_{sum}^*$ \cite{Gan2015Discrete}, which reads
\begin{equation}\label{D42}
D_{sum}^* = \sqrt {(D_2^*)^2+(D_{3,1}^*)^2+(D_3^*)^2+(D_{4,2}^*)^2}.
\end{equation}
In Ref. \cite{Gan2015Discrete}, it has been found that the total non-equilibrium strength $D_{sum}^*$ can be used as a physical criterion to discriminate the two stages of the thermal phase separation. In this work, we aim to show that not only the $D_{sum}^*$ but also each $D_m^*$ can present criteria to distinguish the two stages of phase separation.

In addition, in one of our recent works, we established the relationship between the entropy production rate and the non-equilibrium quantities for multiphase flows, which reads \cite{Zhang-2018arXiv2}
\begin{equation}\label{Eq-Entropy1}
\frac{{d{S_b}}}{{dt}} = \int {\left( {{{\bm{\Delta }}_{3,1}^*}  \cdot \nabla  \frac{1}{T}  - \frac{1}{T} {{\bm{\Delta }}_2^*} :\nabla {\bf{u}}} \right)d{\bf{r}}}.
\end{equation}
There are two source terms that directly contribute to the entropy production including the non-organized energy fluxes (NOEF) and the non-organized momentum fluxes (NOMF). The two terms of entropy production are denoted by $\dot S _{NOEF}$ and $\dot S _{NOMF}$, respectively,
\begin{equation}\label{Eq-Entropy-rate-NOEF}
\dot S _{NOEF} = \int { {{{\bm{\Delta }}_{3,1}^*}  \cdot \nabla  \frac{1}{T} } d{\bf{r}}},
\end{equation}
\begin{equation}\label{Eq-Entropy-rate-NOMF}
\dot S _{NOMF} = \int { { - \frac{1}{T} {{\bm{\Delta }}_2^*} :\nabla {\bf{u}}} d{\bf{r}}},
\end{equation}
where the space integrals are within the whole computational domain. The total entropy production rate is denoted as $\dot S_{sum}$ and it has
\begin{equation}\label{Eq-Entropy-rate-sum}
\dot S _{sum} = \dot S _{NOEF} + \dot S _{NOMF}.
\end{equation}
We find that the $\dot S_{sum}$ increases with time at the SD stage and decreases at the DG stage. The maximum point of $\dot S_{sum}$ indicates the critical time of the phase separation. The profiles of $\dot S_{NOEF}$ and $\dot S_{NOMF}$ possess similar characteristics with $\dot S_{sum}$.

\begin{figure}[h]
\centering
  \includegraphics[height=5cm]{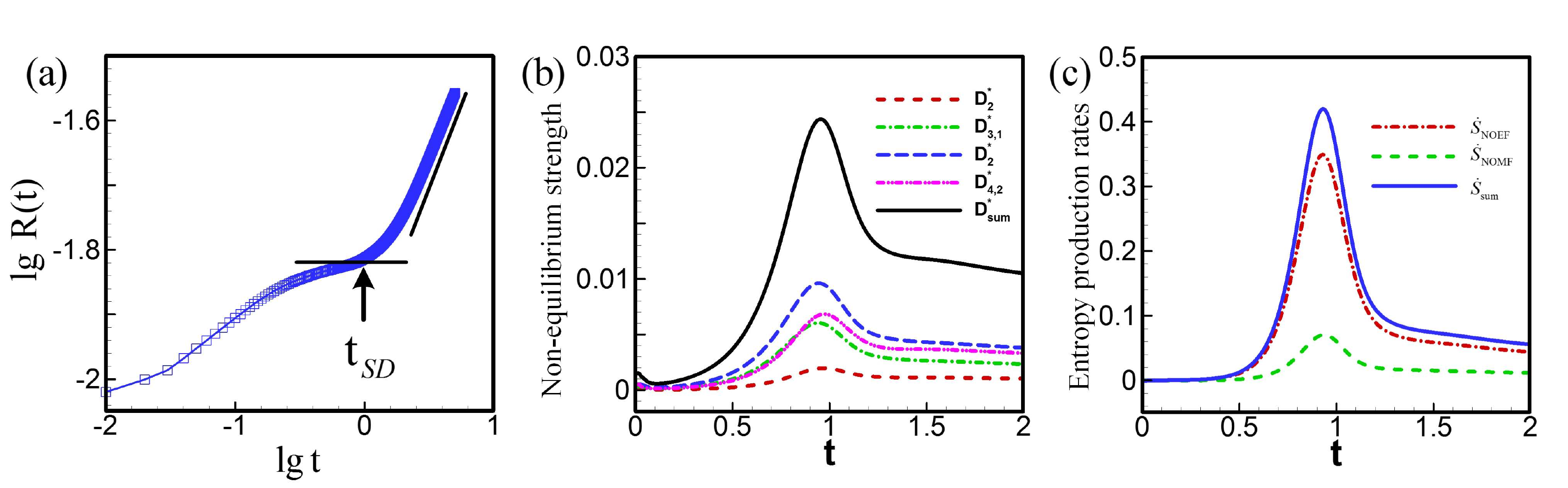}
  \caption{Profiles of the (a) characteristic domain size, (b) non-equilibrium strength, and (c) entropy production rate.}
  \label{fig6}
\end{figure}

The profiles of non-equilibrium strengths and entropy production rates for thermal phase separation are shown in Fig. \ref{fig6}. The characteristic domain size is also plotted in Fig. \ref{fig6} (a) for comparison. The critical time is marked by arrow when the growth of the domain size begins to show a power law. From Fig. \ref{fig6} (b) we can see that the profiles of different non-equilibrium strengths show similar features although their amplitudes are different. They all increase at the first stage and then decrease at the DG stage. The maximum points are a little bit different, but the differences are very small and the maximum points are almost on a vertical line. The profiles of entropy production rates including $\dot S_{NOEF}$, $\dot S_{NOMF}$, and $\dot S_{sum}$ are shown in Fig. \ref{fig6} (c), from which we can see that the maximum points are almost correspond to the same time.

The critical times $t_{SD}$ obtained from different profiles are compared in Tab. \ref{tab-1}. It can not give an accurate value of $t_{SD}$ by means of characteristic domain size, so an approximated range of the critical time is listed in the table. Generally speaking, the values of $t_{SD}$ calculated by non-equilibrium strengths are bigger than those of entropy production rates. The values of $t_{SD}$ provided by the non-equilibrium strengths and entropy production rates are different with each other but the difference is much small. It concludes that those different criteria characterize the phase separation process from different perspective, but they provides the similar features in a statistical sense. The information provided by different criteria are consistent and complementary with each other.

\begin{table}[h]
\caption{Critical times indicated by different criteria.}
\label{tab-1}
\tabcolsep20pt\begin{tabular}{llc}
\hline
   \multicolumn{2}{c}{Different criteria}  & Critical time $t_{SD}$  \\
\hline
Characteristic domain size & $R(t)$ &1.0 $\pm 0.06$ \\
\hline
\multirow{5}{*}{Non-equilibrium strengths} & $D_{2}^*$ & 0.97  \\
                     & $D_{3,1}^*$ & 0.94  \\
                     & $D_{3}^*$ & 0.94  \\
                     & $D_{4,2}^*$ & 0.97 \\
                     & $D^*_{sum}$ & 0.95  \\
  \hline
\multirow{3}{*}{Entropy production rates} & ${\dot S_{NOEF}}$ & 0.93  \\
                     & ${\dot S_{NOMF}}$ & 0.94  \\
                     & ${\dot S_{sum}}$ & 0.93  \\
\hline
\end{tabular}
\end{table}
\section{CONCLUSION}

Discrete Boltzmann modeling presents more convenient physical criteria to discriminate the two stages, spinnodal decomposition and domain growth. These physical criteria,
entropy production rate and various strengths of non-equilibrium, are further investigated and compared with previous rheological and morphological ones which are in fact some geometrical criteria. It is found that the
physical criteria are more efficient to provide critical times for the crossover.
 Each of all those criteria characterizes the phase separation process from its own perspective. The slight differences in critical times obtained via different criteria show the complexity of the phase separation process and are complementary in describing the complex phenomena.

\section{ACKNOWLEDGMENTS}
YZ, AX, and GZ acknowledge the support of National Natural Science Foundation of China (under grant nos. 11475028, 11772064) and  Science Challenge Project (under Grant No. JCKY2016212A501). ZC acknowledges the support of National Natural Science Foundation of China (under grant no. 11502117).


\nocite{*}
\bibliographystyle{aipnum-cp}%
\bibliography{AIP-manuscript}%

\begin{thebibliography}{41}%
\makeatletter
\providecommand \@ifxundefined [1]{%
 \@ifx{#1\undefined}
}%
\providecommand \@ifnum [1]{%
 \ifnum #1\expandafter \@firstoftwo
 \else \expandafter \@secondoftwo
 \fi
}%
\providecommand \@ifx [1]{%
 \ifx #1\expandafter \@firstoftwo
 \else \expandafter \@secondoftwo
 \fi
}%
\providecommand \natexlab [1]{#1}%
\providecommand \enquote  [1]{``#1''}%
\providecommand \bibnamefont  [1]{#1}%
\providecommand \bibfnamefont [1]{#1}%
\providecommand \citenamefont [1]{#1}%
\providecommand \href@noop [0]{\@secondoftwo}%
\providecommand \href [0]{\begingroup \@sanitize@url \@href}%
\providecommand \@href[1]{\@@startlink{#1}\@@href}%
\providecommand \@@href[1]{\endgroup#1\@@endlink}%
\providecommand \@sanitize@url [0]{\catcode `\$12\catcode `\&12\catcode
  `\#12\catcode `\^12\catcode `\_12\catcode `\%12\relax}%
\providecommand \@@startlink[1]{}%
\providecommand \@@endlink[0]{}%
\providecommand \url  [0]{\begingroup\@sanitize@url \@url }%
\providecommand \@url [1]{\endgroup\@href {#1}{\urlprefix }}%
\providecommand \urlprefix  [0]{URL }%
\providecommand \Eprint [0]{\href }%
\providecommand \doibase [0]{http://dx.doi.org/}%
\providecommand \selectlanguage [0]{\@gobble}%
\providecommand \bibinfo  [0]{\@secondoftwo}%
\providecommand \bibfield  [0]{\@secondoftwo}%
\providecommand \translation [1]{[#1]}%
\providecommand \BibitemOpen [0]{}%
\providecommand \bibitemStop [0]{}%
\providecommand \bibitemNoStop [0]{.\EOS\space}%
\providecommand \EOS [0]{\spacefactor3000\relax}%
\providecommand \BibitemShut  [1]{\csname bibitem#1\endcsname}%
\let\auto@bib@innerbib\@empty
\bibitem [{\citenamefont {Sperling}(2006)}]{Sperling2006}%
  \BibitemOpen
  \bibfield  {author} {\bibinfo {author} {\bibfnamefont {L.~H.}\ \bibnamefont
  {Sperling}},\ }\href@noop {} {\emph {\bibinfo {title} {Introduction to
  physical polymer science, Fourth edition}}}\ (\bibinfo  {publisher} {John
  Wiley \& Sons, Inc.,. Hoboken},\ \bibinfo {address} {New Jersey},\ \bibinfo
  {year} {2006})\BibitemShut {NoStop}%
\bibitem [{\citenamefont {Onuki}(2002)}]{PhaseTrans-book}%
  \BibitemOpen
  \bibfield  {author} {\bibinfo {author} {\bibfnamefont {A.}~\bibnamefont
  {Onuki}},\ }\href@noop {} {\emph {\bibinfo {title} {Phase Transition
  Dynamics}}}\ (\bibinfo  {publisher} {Cambridge University Press},\ \bibinfo
  {address} {Cambridge},\ \bibinfo {year} {2002})\ \unskip, pp.\ \bibinfo
  {pages} {47{--}52}\BibitemShut {NoStop}%
\bibitem [{\citenamefont {Xu}, \citenamefont {Zhang},\ and\ \citenamefont
  {Gan}(2016)}]{Xu2015Progress}%
  \BibitemOpen
  \bibfield  {author} {\bibinfo {author} {\bibfnamefont {A.}~\bibnamefont
  {Xu}}, \bibinfo {author} {\bibfnamefont {G.}~\bibnamefont {Zhang}}, \ and\
  \bibinfo {author} {\bibfnamefont {Y.}~\bibnamefont {Gan}},\ }\href@noop {}
  {\bibfield  {journal} {\bibinfo  {journal} {Mechanics in Engineering}\
  }\textbf {\bibinfo {volume} {38}},\ \unskip\ \bibinfo {pages} {361--374}
  (\bibinfo {year} {2016})}\BibitemShut {NoStop}%
\bibitem [{\citenamefont {Gan}\ \emph {et~al.}(2015)\citenamefont {Gan},
  \citenamefont {Xu}, \citenamefont {Zhang},\ and\ \citenamefont
  {Succi}}]{Gan2015Discrete}%
  \BibitemOpen
  \bibfield  {author} {\bibinfo {author} {\bibfnamefont {Y.}~\bibnamefont
  {Gan}}, \bibinfo {author} {\bibfnamefont {A.}~\bibnamefont {Xu}}, \bibinfo
  {author} {\bibfnamefont {G.}~\bibnamefont {Zhang}}, \ and\ \bibinfo {author}
  {\bibfnamefont {S.}~\bibnamefont {Succi}},\ }\href@noop {} {\bibfield
  {journal} {\bibinfo  {journal} {Soft Matter}\ }\textbf {\bibinfo {volume}
  {11}},\ \unskip\ \bibinfo {pages} {5336--5345} (\bibinfo {year}
  {2015})}\BibitemShut {NoStop}%
\bibitem [{\citenamefont {Ye}\ \emph {et~al.}(2013)\citenamefont {Ye},
  \citenamefont {Lin}, \citenamefont {Huang}, \citenamefont {Liang},\ and\
  \citenamefont {Xu}}]{Ye2013Polymer}%
  \BibitemOpen
  \bibfield  {author} {\bibinfo {author} {\bibfnamefont {X.~Y.}\ \bibnamefont
  {Ye}}, \bibinfo {author} {\bibfnamefont {F.~W.}\ \bibnamefont {Lin}},
  \bibinfo {author} {\bibfnamefont {X.~J.}\ \bibnamefont {Huang}}, \bibinfo
  {author} {\bibfnamefont {H.~Q.}\ \bibnamefont {Liang}}, \ and\ \bibinfo
  {author} {\bibfnamefont {Z.~K.}\ \bibnamefont {Xu}},\ }\href@noop {}
  {\bibfield  {journal} {\bibinfo  {journal} {RSC Adv.}\ }\textbf {\bibinfo
  {volume} {3}},\ \unskip\ \bibinfo {pages} {13851--13858} (\bibinfo {year}
  {2013})}\BibitemShut {NoStop}%
\bibitem [{\citenamefont {Yeganeh}, \citenamefont {Goharpey},\ and\
  \citenamefont {Foudazi}(2014)}]{Yeganeh2014Anomalous}%
  \BibitemOpen
  \bibfield  {author} {\bibinfo {author} {\bibfnamefont {J.}~\bibnamefont
  {Yeganeh}}, \bibinfo {author} {\bibfnamefont {F.}~\bibnamefont {Goharpey}}, \
  and\ \bibinfo {author} {\bibfnamefont {R.}~\bibnamefont {Foudazi}},\
  }\href@noop {} {\bibfield  {journal} {\bibinfo  {journal} {RSC Adv.}\
  }\textbf {\bibinfo {volume} {4}},\ \unskip\ \bibinfo {pages} {12809--12825}
  (\bibinfo {year} {2014})}\BibitemShut {NoStop}%
\bibitem [{\citenamefont {Iwashita}\ and\ \citenamefont
  {Tanaka}(2006)}]{Iwashita2006Self}%
  \BibitemOpen
  \bibfield  {author} {\bibinfo {author} {\bibfnamefont {Y.}~\bibnamefont
  {Iwashita}}\ and\ \bibinfo {author} {\bibfnamefont {H.}~\bibnamefont
  {Tanaka}},\ }\href@noop {} {\bibfield  {journal} {\bibinfo  {journal} {Nat.
  Mater.}\ }\textbf {\bibinfo {volume} {5}},\ \unskip\ \bibinfo {pages}
  {147--52} (\bibinfo {year} {2006})}\BibitemShut {NoStop}%
\bibitem [{\citenamefont {Succi}(2018)}]{Succi2018The}%
  \BibitemOpen
  \bibfield  {author} {\bibinfo {author} {\bibfnamefont {S.}~\bibnamefont
  {Succi}},\ }\href@noop {} {\emph {\bibinfo {title} {The Lattice Boltzmann
  Equation: For Complex States of Flowing Matter}}}\ (\bibinfo  {publisher}
  {Oxford univeristy press},\ \bibinfo {address} {Oxford},\ \bibinfo {year}
  {2018})\BibitemShut {NoStop}%
\bibitem [{\citenamefont {Huang}, \citenamefont {Sukop},\ and\ \citenamefont
  {Lu}(2015)}]{Huang2015Multiphase}%
  \BibitemOpen
  \bibfield  {author} {\bibinfo {author} {\bibfnamefont {H.}~\bibnamefont
  {Huang}}, \bibinfo {author} {\bibfnamefont {M.~C.}\ \bibnamefont {Sukop}}, \
  and\ \bibinfo {author} {\bibfnamefont {X.~Y.}\ \bibnamefont {Lu}},\
  }\href@noop {} {\emph {\bibinfo {title} {Multiphase Lattice Boltzmann
  Methods: Theory and Application}}}\ (\bibinfo  {publisher} {John Wiley \&
  Sons, Ltd},\ \bibinfo {address} {West Sussex},\ \bibinfo {year}
  {2015})\BibitemShut {NoStop}%
\bibitem [{\citenamefont {Gunstensen}\ \emph {et~al.}(1991)\citenamefont
  {Gunstensen}, \citenamefont {Rothman}, \citenamefont {Zaleski},\ and\
  \citenamefont {Zanetti}}]{Gunstensen1991Lattice}%
  \BibitemOpen
  \bibfield  {author} {\bibinfo {author} {\bibfnamefont {A.~K.}\ \bibnamefont
  {Gunstensen}}, \bibinfo {author} {\bibfnamefont {D.~H.}\ \bibnamefont
  {Rothman}}, \bibinfo {author} {\bibfnamefont {S.}~\bibnamefont {Zaleski}}, \
  and\ \bibinfo {author} {\bibfnamefont {G.}~\bibnamefont {Zanetti}},\
  }\href@noop {} {\bibfield  {journal} {\bibinfo  {journal} {Phys. Rev. A}\
  }\textbf {\bibinfo {volume} {43}},\ \unskip\ \bibinfo {pages} {4320--4327}
  (\bibinfo {year} {1991})}\BibitemShut {NoStop}%
\bibitem [{\citenamefont {Swift}\ \emph {et~al.}(1996)\citenamefont {Swift},
  \citenamefont {Orlandini}, \citenamefont {Osborn},\ and\ \citenamefont
  {Yeomans}}]{Swift1996Lattice}%
  \BibitemOpen
  \bibfield  {author} {\bibinfo {author} {\bibfnamefont {M.~R.}\ \bibnamefont
  {Swift}}, \bibinfo {author} {\bibfnamefont {E.}~\bibnamefont {Orlandini}},
  \bibinfo {author} {\bibfnamefont {W.~R.}\ \bibnamefont {Osborn}}, \ and\
  \bibinfo {author} {\bibfnamefont {J.~M.}\ \bibnamefont {Yeomans}},\
  }\href@noop {} {\bibfield  {journal} {\bibinfo  {journal} {Phys. Rev. E:
  Stat. Phys., Plasmas, Fluids, Relat. Interdiscip. Top.}\ }\textbf {\bibinfo
  {volume} {54}},\ p.\ \bibinfo {pages} {5041} (\bibinfo {year}
  {1996})}\BibitemShut {NoStop}%
\bibitem [{\citenamefont {Falcucci}, \citenamefont {Ubertini},\ and\
  \citenamefont {Succi}(2010)}]{Falcucci2010Lattice}%
  \BibitemOpen
  \bibfield  {author} {\bibinfo {author} {\bibfnamefont {G.}~\bibnamefont
  {Falcucci}}, \bibinfo {author} {\bibfnamefont {S.}~\bibnamefont {Ubertini}},
  \ and\ \bibinfo {author} {\bibfnamefont {S.}~\bibnamefont {Succi}},\
  }\href@noop {} {\bibfield  {journal} {\bibinfo  {journal} {Soft Matter}\
  }\textbf {\bibinfo {volume} {6}},\ \unskip\ \bibinfo {pages} {4357--4365}
  (\bibinfo {year} {2010})}\BibitemShut {NoStop}%
\bibitem [{\citenamefont {Shan}\ and\ \citenamefont
  {Chen}(1993)}]{Shan1993Lattice}%
  \BibitemOpen
  \bibfield  {author} {\bibinfo {author} {\bibfnamefont {X.}~\bibnamefont
  {Shan}}\ and\ \bibinfo {author} {\bibfnamefont {H.}~\bibnamefont {Chen}},\
  }\href@noop {} {\bibfield  {journal} {\bibinfo  {journal} {Phys. Rev. E:
  Stat. Phys., Plasmas, Fluids, Relat. Interdiscip. Top.}\ }\textbf {\bibinfo
  {volume} {47}},\ \unskip\ \bibinfo {pages} {1815--1819} (\bibinfo {year}
  {1993})}\BibitemShut {NoStop}%
\bibitem [{\citenamefont {Shan}\ and\ \citenamefont
  {Chen}(1994)}]{Shan1994Simulation}%
  \BibitemOpen
  \bibfield  {author} {\bibinfo {author} {\bibfnamefont {X.}~\bibnamefont
  {Shan}}\ and\ \bibinfo {author} {\bibfnamefont {H.}~\bibnamefont {Chen}},\
  }\href@noop {} {\bibfield  {journal} {\bibinfo  {journal} {Phys. Rev. E:
  Stat. Phys., Plasmas, Fluids, Relat. Interdiscip. Top.}\ }\textbf {\bibinfo
  {volume} {49}},\ \unskip\ \bibinfo {pages} {2941--2948} (\bibinfo {year}
  {1994})}\BibitemShut {NoStop}%
\bibitem [{\citenamefont {Swift}, \citenamefont {Osborn},\ and\ \citenamefont
  {Yeomans}(1995)}]{Swift1995Lattice}%
  \BibitemOpen
  \bibfield  {author} {\bibinfo {author} {\bibfnamefont {M.~R.}\ \bibnamefont
  {Swift}}, \bibinfo {author} {\bibfnamefont {W.~R.}\ \bibnamefont {Osborn}}, \
  and\ \bibinfo {author} {\bibfnamefont {J.~M.}\ \bibnamefont {Yeomans}},\
  }\href@noop {} {\bibfield  {journal} {\bibinfo  {journal} {Phys. Rev. Lett.}\
  }\textbf {\bibinfo {volume} {75}},\ \unskip\ \bibinfo {pages} {830--833}
  (\bibinfo {year} {1995})}\BibitemShut {NoStop}%
\bibitem [{\citenamefont {He}, \citenamefont {Chen},\ and\ \citenamefont
  {Zhang}(1999)}]{He1999A}%
  \BibitemOpen
  \bibfield  {author} {\bibinfo {author} {\bibfnamefont {X.}~\bibnamefont
  {He}}, \bibinfo {author} {\bibfnamefont {S.}~\bibnamefont {Chen}}, \ and\
  \bibinfo {author} {\bibfnamefont {R.}~\bibnamefont {Zhang}},\ }\href@noop {}
  {\bibfield  {journal} {\bibinfo  {journal} {J. Comput. Phys.}\ }\textbf
  {\bibinfo {volume} {152}},\ \unskip\ \bibinfo {pages} {642--663} (\bibinfo
  {year} {1999})}\BibitemShut {NoStop}%
\bibitem [{\citenamefont {Li}\ \emph {et~al.}(2016)\citenamefont {Li},
  \citenamefont {Luo}, \citenamefont {Kang}, \citenamefont {He}, \citenamefont
  {Chen},\ and\ \citenamefont {Liu}}]{Li2016Lattice}%
  \BibitemOpen
  \bibfield  {author} {\bibinfo {author} {\bibfnamefont {Q.}~\bibnamefont
  {Li}}, \bibinfo {author} {\bibfnamefont {K.~H.}\ \bibnamefont {Luo}},
  \bibinfo {author} {\bibfnamefont {Q.~J.}\ \bibnamefont {Kang}}, \bibinfo
  {author} {\bibfnamefont {Y.~L.}\ \bibnamefont {He}}, \bibinfo {author}
  {\bibfnamefont {Q.}~\bibnamefont {Chen}}, \ and\ \bibinfo {author}
  {\bibfnamefont {Q.}~\bibnamefont {Liu}},\ }\href@noop {} {\bibfield
  {journal} {\bibinfo  {journal} {Prog. Energy Combust. Sci.}\ }\textbf
  {\bibinfo {volume} {52}},\ \unskip\ \bibinfo {pages} {62--105} (\bibinfo
  {year} {2016})}\BibitemShut {NoStop}%
\bibitem [{\citenamefont {Bray}(1995)}]{A1995Theory}%
  \BibitemOpen
  \bibfield  {author} {\bibinfo {author} {\bibfnamefont {A.}~\bibnamefont
  {Bray}},\ }\href@noop {} {\bibfield  {journal} {\bibinfo  {journal} {Phys.
  A}\ }\textbf {\bibinfo {volume} {194}},\ \unskip\ \bibinfo {pages} {41--52}
  (\bibinfo {year} {1995})}\BibitemShut {NoStop}%
\bibitem [{\citenamefont {Xu}, \citenamefont {Gonnella},\ and\ \citenamefont
  {Lamura}(2006)}]{Xu2006Morphologies}%
  \BibitemOpen
  \bibfield  {author} {\bibinfo {author} {\bibfnamefont {A.}~\bibnamefont
  {Xu}}, \bibinfo {author} {\bibfnamefont {G.}~\bibnamefont {Gonnella}}, \ and\
  \bibinfo {author} {\bibfnamefont {A.}~\bibnamefont {Lamura}},\ }\href@noop {}
  {\bibfield  {journal} {\bibinfo  {journal} {Phys. Rev. E: Stat., Nonlinear,
  Soft Matter Phys.}\ }\textbf {\bibinfo {volume} {74}},\ p.\ \bibinfo {pages}
  {011505} (\bibinfo {year} {2006})}\BibitemShut {NoStop}%
\bibitem [{\citenamefont {Allen}\ and\ \citenamefont
  {Cahn}(1976)}]{Allen1976Mechanisms}%
  \BibitemOpen
  \bibfield  {author} {\bibinfo {author} {\bibfnamefont {S.~M.}\ \bibnamefont
  {Allen}}\ and\ \bibinfo {author} {\bibfnamefont {J.~W.}\ \bibnamefont
  {Cahn}},\ }\href@noop {} {\bibfield  {journal} {\bibinfo  {journal} {Acta
  Metall.}\ }\textbf {\bibinfo {volume} {24}},\ \unskip\ \bibinfo {pages}
  {425--437} (\bibinfo {year} {1976})}\BibitemShut {NoStop}%
\bibitem [{\citenamefont {Gan}\ \emph {et~al.}(2011)\citenamefont {Gan},
  \citenamefont {Xu}, \citenamefont {Zhang}, \citenamefont {Li},\ and\
  \citenamefont {Li}}]{Gan2011Phase}%
  \BibitemOpen
  \bibfield  {author} {\bibinfo {author} {\bibfnamefont {Y.}~\bibnamefont
  {Gan}}, \bibinfo {author} {\bibfnamefont {A.}~\bibnamefont {Xu}}, \bibinfo
  {author} {\bibfnamefont {G.}~\bibnamefont {Zhang}}, \bibinfo {author}
  {\bibfnamefont {Y.}~\bibnamefont {Li}}, \ and\ \bibinfo {author}
  {\bibfnamefont {H.}~\bibnamefont {Li}},\ }\href@noop {} {\bibfield  {journal}
  {\bibinfo  {journal} {Phys. Rev. E: Stat., Nonlinear, Soft Matter Phys.}\
  }\textbf {\bibinfo {volume} {84}},\ p.\ \bibinfo {pages} {046715} (\bibinfo
  {year} {2011})}\BibitemShut {NoStop}%
\bibitem [{\citenamefont {Zhang}\ \emph
  {et~al.}(2018{\natexlab{a}})\citenamefont {Zhang}, \citenamefont {Xu},
  \citenamefont {Zhang}, \citenamefont {Gan}, \citenamefont {Chen},\ and\
  \citenamefont {Succi}}]{Zhang-2018arXiv2}%
  \BibitemOpen
  \bibfield  {author} {\bibinfo {author} {\bibfnamefont {Y.}~\bibnamefont
  {Zhang}}, \bibinfo {author} {\bibfnamefont {A.}~\bibnamefont {Xu}}, \bibinfo
  {author} {\bibfnamefont {G.}~\bibnamefont {Zhang}}, \bibinfo {author}
  {\bibfnamefont {Y.}~\bibnamefont {Gan}}, \bibinfo {author} {\bibfnamefont
  {Z.}~\bibnamefont {Chen}}, \ and\ \bibinfo {author} {\bibfnamefont
  {S.}~\bibnamefont {Succi}},\ }\href@noop {} {\bibfield  {journal} {\bibinfo
  {journal} {arXiv:1808.07698}\ } (\bibinfo {year}
  {2018}{\natexlab{a}})}\BibitemShut {NoStop}%
\bibitem [{\citenamefont {Gan}\ \emph {et~al.}(2018)\citenamefont {Gan},
  \citenamefont {Xu}, \citenamefont {Zhang},\ and\ \citenamefont
  {Lai}}]{Gan2017Three}%
  \BibitemOpen
  \bibfield  {author} {\bibinfo {author} {\bibfnamefont {Y.}~\bibnamefont
  {Gan}}, \bibinfo {author} {\bibfnamefont {A.}~\bibnamefont {Xu}}, \bibinfo
  {author} {\bibfnamefont {G.}~\bibnamefont {Zhang}}, \ and\ \bibinfo {author}
  {\bibfnamefont {H.}~\bibnamefont {Lai}},\ }\href@noop {} {\bibfield
  {journal} {\bibinfo  {journal} {Proc. Inst. Mech. Eng., Part C}\ }\textbf
  {\bibinfo {volume} {232}},\ \unskip\ \bibinfo {pages} {477--490} (\bibinfo
  {year} {2018})}\BibitemShut {NoStop}%
\bibitem [{\citenamefont {Chen}, \citenamefont {Xu},\ and\ \citenamefont
  {Zhang}(2016)}]{Feng2016Viscosity}%
  \BibitemOpen
  \bibfield  {author} {\bibinfo {author} {\bibfnamefont {F.}~\bibnamefont
  {Chen}}, \bibinfo {author} {\bibfnamefont {A.}~\bibnamefont {Xu}}, \ and\
  \bibinfo {author} {\bibfnamefont {G.}~\bibnamefont {Zhang}},\ }\href@noop {}
  {\bibfield  {journal} {\bibinfo  {journal} {Front. Phys.}\ }\textbf {\bibinfo
  {volume} {11}},\ \unskip\ \bibinfo {pages} {183--196} (\bibinfo {year}
  {2016})}\BibitemShut {NoStop}%
\bibitem [{\citenamefont {Lai}\ \emph {et~al.}(2016)\citenamefont {Lai},
  \citenamefont {Xu}, \citenamefont {Zhang}, \citenamefont {Gan}, \citenamefont
  {Ying},\ and\ \citenamefont {Succi}}]{Lai2016Nonequilibrium}%
  \BibitemOpen
  \bibfield  {author} {\bibinfo {author} {\bibfnamefont {H.}~\bibnamefont
  {Lai}}, \bibinfo {author} {\bibfnamefont {A.}~\bibnamefont {Xu}}, \bibinfo
  {author} {\bibfnamefont {G.}~\bibnamefont {Zhang}}, \bibinfo {author}
  {\bibfnamefont {Y.}~\bibnamefont {Gan}}, \bibinfo {author} {\bibfnamefont
  {Y.}~\bibnamefont {Ying}}, \ and\ \bibinfo {author} {\bibfnamefont
  {S.}~\bibnamefont {Succi}},\ }\href@noop {} {\bibfield  {journal} {\bibinfo
  {journal} {Phys. Rev. E}\ }\textbf {\bibinfo {volume} {94}},\ p.\ \bibinfo
  {pages} {023106} (\bibinfo {year} {2016})}\BibitemShut {NoStop}%
\bibitem [{\citenamefont {Lin}\ \emph {et~al.}(2017{\natexlab{a}})\citenamefont
  {Lin}, \citenamefont {Xu}, \citenamefont {Zhang}, \citenamefont {Luo},\ and\
  \citenamefont {Li}}]{Lin2017Discrete}%
  \BibitemOpen
  \bibfield  {author} {\bibinfo {author} {\bibfnamefont {C.}~\bibnamefont
  {Lin}}, \bibinfo {author} {\bibfnamefont {A.}~\bibnamefont {Xu}}, \bibinfo
  {author} {\bibfnamefont {G.}~\bibnamefont {Zhang}}, \bibinfo {author}
  {\bibfnamefont {K.~H.}\ \bibnamefont {Luo}}, \ and\ \bibinfo {author}
  {\bibfnamefont {Y.}~\bibnamefont {Li}},\ }\href@noop {} {\bibfield  {journal}
  {\bibinfo  {journal} {Physical Review E}\ }\textbf {\bibinfo {volume} {96}},\
  p.\ \bibinfo {pages} {053305} (\bibinfo {year}
  {2017}{\natexlab{a}})}\BibitemShut {NoStop}%
\bibitem [{\citenamefont {Lin}\ \emph {et~al.}(2017{\natexlab{b}})\citenamefont
  {Lin}, \citenamefont {Luo}, \citenamefont {Fei},\ and\ \citenamefont
  {Succi}}]{Lin2017A}%
  \BibitemOpen
  \bibfield  {author} {\bibinfo {author} {\bibfnamefont {C.}~\bibnamefont
  {Lin}}, \bibinfo {author} {\bibfnamefont {K.~H.}\ \bibnamefont {Luo}},
  \bibinfo {author} {\bibfnamefont {L.}~\bibnamefont {Fei}}, \ and\ \bibinfo
  {author} {\bibfnamefont {S.}~\bibnamefont {Succi}},\ }\href@noop {}
  {\bibfield  {journal} {\bibinfo  {journal} {Sci. Rep.}\ }\textbf {\bibinfo
  {volume} {7}},\ p.\ \bibinfo {pages} {14580} (\bibinfo {year}
  {2017}{\natexlab{b}})}\BibitemShut {NoStop}%
\bibitem [{\citenamefont {Zhang}\ \emph
  {et~al.}(2018{\natexlab{b}})\citenamefont {Zhang}, \citenamefont {Xu},
  \citenamefont {Zhang}, \citenamefont {Chen},\ and\ \citenamefont
  {Wang}}]{ZYD-2018arXiv}%
  \BibitemOpen
  \bibfield  {author} {\bibinfo {author} {\bibfnamefont {Y.}~\bibnamefont
  {Zhang}}, \bibinfo {author} {\bibfnamefont {A.}~\bibnamefont {Xu}}, \bibinfo
  {author} {\bibfnamefont {G.}~\bibnamefont {Zhang}}, \bibinfo {author}
  {\bibfnamefont {Z.}~\bibnamefont {Chen}}, \ and\ \bibinfo {author}
  {\bibfnamefont {P.}~\bibnamefont {Wang}},\ }\href@noop {} {\bibfield
  {journal} {\bibinfo  {journal} {arXiv:1801.02649}\ } (\bibinfo {year}
  {2018}{\natexlab{b}})}\BibitemShut {NoStop}%
\bibitem [{\citenamefont {Watari}\ and\ \citenamefont
  {Tsutahara}(2003)}]{Watari2003Two}%
  \BibitemOpen
  \bibfield  {author} {\bibinfo {author} {\bibfnamefont {M.}~\bibnamefont
  {Watari}}\ and\ \bibinfo {author} {\bibfnamefont {M.}~\bibnamefont
  {Tsutahara}},\ }\href@noop {} {\bibfield  {journal} {\bibinfo  {journal}
  {Phys. Rev. E: Stat., Nonlinear, Soft Matter Phys.}\ }\textbf {\bibinfo
  {volume} {67}},\ p.\ \bibinfo {pages} {036306} (\bibinfo {year}
  {2003})}\BibitemShut {NoStop}%
\bibitem [{\citenamefont {Watari}(2016)}]{Watari2016Is}%
  \BibitemOpen
  \bibfield  {author} {\bibinfo {author} {\bibfnamefont {M.}~\bibnamefont
  {Watari}},\ }\href@noop {} {\bibfield  {journal} {\bibinfo  {journal} {J.
  Fluids Eng.}\ }\textbf {\bibinfo {volume} {138}} (\bibinfo {year}
  {2016})}\BibitemShut {NoStop}%
\bibitem [{\citenamefont {Gonnella}, \citenamefont {Lamura},\ and\
  \citenamefont {Sofonea}(2007)}]{Gonnella2007Lattice}%
  \BibitemOpen
  \bibfield  {author} {\bibinfo {author} {\bibfnamefont {G.}~\bibnamefont
  {Gonnella}}, \bibinfo {author} {\bibfnamefont {A.}~\bibnamefont {Lamura}}, \
  and\ \bibinfo {author} {\bibfnamefont {V.}~\bibnamefont {Sofonea}},\
  }\href@noop {} {\bibfield  {journal} {\bibinfo  {journal} {Phys. Rev. E:
  Stat., Nonlinear, Soft Matter Phys.}\ }\textbf {\bibinfo {volume} {76}},\ p.\
  \bibinfo {pages} {036703} (\bibinfo {year} {2007})}\BibitemShut {NoStop}%
\bibitem [{\citenamefont {Klimontovich}(1982)}]{Klimontovich-book}%
  \BibitemOpen
  \bibfield  {author} {\bibinfo {author} {\bibfnamefont {Y.}~\bibnamefont
  {Klimontovich}},\ }\href@noop {} {\emph {\bibinfo {title} {Kinetic Theory of
  Nonideal Gases and Nonideal Plasmas}}}\ (\bibinfo  {publisher} {Pergamon},\
  \bibinfo {address} {Oxford},\ \bibinfo {year} {1982})\BibitemShut {NoStop}%
\bibitem [{\citenamefont {Onuki}(2005)}]{Onuki2005Dynamic}%
  \BibitemOpen
  \bibfield  {author} {\bibinfo {author} {\bibfnamefont {A.}~\bibnamefont
  {Onuki}},\ }\href@noop {} {\bibfield  {journal} {\bibinfo  {journal} {Phys.
  Rev. Lett.}\ }\textbf {\bibinfo {volume} {94}},\ p.\ \bibinfo {pages}
  {054501} (\bibinfo {year} {2005})}\BibitemShut {NoStop}%
\bibitem [{\citenamefont {Onuki}(2007)}]{Onuki2007Dynamic}%
  \BibitemOpen
  \bibfield  {author} {\bibinfo {author} {\bibfnamefont {A.}~\bibnamefont
  {Onuki}},\ }\href@noop {} {\bibfield  {journal} {\bibinfo  {journal} {Phys.
  Rev. E: Stat., Nonlinear, Soft Matter Phys.}\ }\textbf {\bibinfo {volume}
  {75}},\ p.\ \bibinfo {pages} {036304} (\bibinfo {year} {2007})}\BibitemShut
  {NoStop}%
\bibitem [{\citenamefont {Tiribocchi}\ \emph {et~al.}(2009)\citenamefont
  {Tiribocchi}, \citenamefont {Stella}, \citenamefont {Gonnella},\ and\
  \citenamefont {Lamura}}]{Tiribocchi2009Hybrid}%
  \BibitemOpen
  \bibfield  {author} {\bibinfo {author} {\bibfnamefont {A.}~\bibnamefont
  {Tiribocchi}}, \bibinfo {author} {\bibfnamefont {N.}~\bibnamefont {Stella}},
  \bibinfo {author} {\bibfnamefont {G.}~\bibnamefont {Gonnella}}, \ and\
  \bibinfo {author} {\bibfnamefont {A.}~\bibnamefont {Lamura}},\ }\href@noop {}
  {\bibfield  {journal} {\bibinfo  {journal} {Phys. Rev. E: Stat., Nonlinear,
  Soft Matter Phys.}\ }\textbf {\bibinfo {volume} {80}},\ p.\ \bibinfo {pages}
  {026701} (\bibinfo {year} {2009})}\BibitemShut {NoStop}%
\bibitem [{\citenamefont {Gan}\ \emph {et~al.}(2012)\citenamefont {Gan},
  \citenamefont {Xu}, \citenamefont {Zhang},\ and\ \citenamefont
  {Li}}]{Gan2012FFT}%
  \BibitemOpen
  \bibfield  {author} {\bibinfo {author} {\bibfnamefont {Y.}~\bibnamefont
  {Gan}}, \bibinfo {author} {\bibfnamefont {A.}~\bibnamefont {Xu}}, \bibinfo
  {author} {\bibfnamefont {G.}~\bibnamefont {Zhang}}, \ and\ \bibinfo {author}
  {\bibfnamefont {Y.}~\bibnamefont {Li}},\ }\href@noop {} {\bibfield  {journal}
  {\bibinfo  {journal} {Commun. Theor. Phys.}\ }\textbf {\bibinfo {volume}
  {57}},\ \unskip\ \bibinfo {pages} {681--694} (\bibinfo {year}
  {2012})}\BibitemShut {NoStop}%
\bibitem [{\citenamefont {Mecke}(1996)}]{Mecke1996Morphological}%
  \BibitemOpen
  \bibfield  {author} {\bibinfo {author} {\bibfnamefont {K.~R.}\ \bibnamefont
  {Mecke}},\ }\href@noop {} {\bibfield  {journal} {\bibinfo  {journal} {Phys.
  Rev. E: Stat. Phys., Plasmas, Fluids, Relat. Interdiscip. Top.}\ }\textbf
  {\bibinfo {volume} {53}},\ \unskip\ \bibinfo {pages} {4794--4800} (\bibinfo
  {year} {1996})}\BibitemShut {NoStop}%
\bibitem [{\citenamefont {Xu}\ \emph {et~al.}(2009)\citenamefont {Xu},
  \citenamefont {Zhang}, \citenamefont {Pan}, \citenamefont {Zhang},\ and\
  \citenamefont {Zhu}}]{Xu2009Morphological}%
  \BibitemOpen
  \bibfield  {author} {\bibinfo {author} {\bibfnamefont {A.}~\bibnamefont
  {Xu}}, \bibinfo {author} {\bibfnamefont {G.}~\bibnamefont {Zhang}}, \bibinfo
  {author} {\bibfnamefont {X.~F.}\ \bibnamefont {Pan}}, \bibinfo {author}
  {\bibfnamefont {P.}~\bibnamefont {Zhang}}, \ and\ \bibinfo {author}
  {\bibfnamefont {J.}~\bibnamefont {Zhu}},\ }\href@noop {} {\bibfield
  {journal} {\bibinfo  {journal} {J. Phys. D: Appl. Phys.}\ }\textbf {\bibinfo
  {volume} {42}} (\bibinfo {year} {2009})}\BibitemShut {NoStop}%
\bibitem [{\citenamefont {Xu}\ \emph {et~al.}(2016)\citenamefont {Xu},
  \citenamefont {Zhang}, \citenamefont {Ying},\ and\ \citenamefont
  {Wang}}]{Xu2016Complex}%
  \BibitemOpen
  \bibfield  {author} {\bibinfo {author} {\bibfnamefont {A.}~\bibnamefont
  {Xu}}, \bibinfo {author} {\bibfnamefont {G.}~\bibnamefont {Zhang}}, \bibinfo
  {author} {\bibfnamefont {Y.}~\bibnamefont {Ying}}, \ and\ \bibinfo {author}
  {\bibfnamefont {C.}~\bibnamefont {Wang}},\ }\href@noop {} {\bibfield
  {journal} {\bibinfo  {journal} {Sci. China (Phys.,Mech. Astron.)}\ }\textbf
  {\bibinfo {volume} {59}},\ p.\ \bibinfo {pages} {650501} (\bibinfo {year}
  {2016})}\BibitemShut {NoStop}%
\bibitem [{\citenamefont {Sofonea}\ and\ \citenamefont
  {Mecke}(1999)}]{Sofonea1999Morphological}%
  \BibitemOpen
  \bibfield  {author} {\bibinfo {author} {\bibfnamefont {V.}~\bibnamefont
  {Sofonea}}\ and\ \bibinfo {author} {\bibfnamefont {K.~R.}\ \bibnamefont
  {Mecke}},\ }\href@noop {} {\bibfield  {journal} {\bibinfo  {journal} {Eur.
  Phys. J. B}\ }\textbf {\bibinfo {volume} {8}},\ \unskip\ \bibinfo {pages}
  {99--112} (\bibinfo {year} {1999})}\BibitemShut {NoStop}%
\bibitem [{\citenamefont {Xu}\ \emph {et~al.}(2012)\citenamefont {Xu},
  \citenamefont {Zhang}, \citenamefont {Gan}, \citenamefont {Chen},\ and\
  \citenamefont {Yu}}]{Xu2012Lattice}%
  \BibitemOpen
  \bibfield  {author} {\bibinfo {author} {\bibfnamefont {A.}~\bibnamefont
  {Xu}}, \bibinfo {author} {\bibfnamefont {G.}~\bibnamefont {Zhang}}, \bibinfo
  {author} {\bibfnamefont {Y.}~\bibnamefont {Gan}}, \bibinfo {author}
  {\bibfnamefont {F.}~\bibnamefont {Chen}}, \ and\ \bibinfo {author}
  {\bibfnamefont {X.}~\bibnamefont {Yu}},\ }\href@noop {} {\bibfield  {journal}
  {\bibinfo  {journal} {Front. Phys.}\ }\textbf {\bibinfo {volume} {7}},\
  \unskip\ \bibinfo {pages} {582--600} (\bibinfo {year} {2012})}\BibitemShut
  {NoStop}%
\end{thebibliography}%

\end{document}